# A micro-structured continuum modelling compacting fluid-saturated grounds: the effects of pore-size scale parameter

F. dell'Isola and L. Rosa, Rome, Italy, and C. Woźniak, Warsaw, Poland

**Summary.** The effect of a "pore-size" length-scale parameter $l$ on compaction of grounds with fluid inclusions is studied. They are modelled as continua endowed with micro-structure by means of the macro-modelling procedure proposed in [2]. We show the dependence of field evolution equations on the micro-structure parameter $l$ and compare our model with the homogenized asymptotic ones. The consideration of the pore size $l$ allows us to forecast the onset of *micro-displacement* waves as a consequence of a ground settling and to suggest a possible description of the genesis of certain microearthquakes [5] [6].

## 1 Introduction

In [1] a new way is suggested to study the effect of the length scale parameter — characterizing the size of the pores — in the mechanical interaction of a fluid with a porous matrix.

The method used there (called in [2] macro-modelling procedure) is very powerful and relatively easy to apply. Here we want to use it to characterize the dynamical evolution of a system $\mathscr{S}$ constituted by a porous solid visco-elastic matrix with non connected periodically distributed inclusions saturated by a viscous fluid. It is well known that the study of this system is very difficult from the mathematical point of view if it is modelled in a *refined* way: i.e. by means of two distinct 3-D Cauchy continua occupying, in every configuration, two disjoint but geometrically *tightly nested* regions of space.

Indeed the complex geometrical structure of the system leads to a coupled (by suitable boundary conditions) system of partial differential equations (PDE) for the displacement field of the solid matrix and for the velocity field of the included fluid. To overcome these difficulties different methods are available in the literature. Here we will mention three of them:

(i) In [7] an asymptotic analysis is proposed which leads to the introduction of homogenized continua as models for considered periodically inhomogeneous systems. These continua are characterized in terms of effective coefficients appearing in their constitutive equations. These effective coefficients are found by means of a complex mathematical procedure involving the solution of a boundary value problem for a prototype PDE in the periodicity cell.

(ii) In [16] — instead — a direct approach is considered. The kinematics of the substantial point of the continuum is described in terms of some descriptors of the micro state of the system. In order to find the evolution equations for these micro-state descriptors some balance equations have to be postulated in every considered instance.

(iii) In [1], [3], [8], [9], although using different mathematical tools, a kind of compromise between the first and second approach is suggested. Refraining from a detailed description of the displacement fields at the micro-level, the evolution equations for the micro-state descriptors are deduced by assuming that they determine with a certain approximation the micro-fields introduced in the *refined 3-D* description. This deduction is possible by the following two steps:

a) First, suitable *micro shape functions* are chosen (respecting the criteria found in [2], [3]).
b) Second, an expression for the micro-fields in terms of micro shape functions, *the micro-descriptors* and the *macro-displacements* is used in the weak form of force balance. As an important by-product macro constitutive equations in terms of micro shape functions and micro constitutive equations are determined.

The first two methods show some drawbacks:

(i) In the method mentioned in (i) the solution of the prototype PDE in most cases is very difficult to construct because of the geometry of the periodicity cell and is solved, in general, only by means of numerical methods. Finally, as homogenized asymptotic models are obtained with a limiting procedure for vanishing periodicity cell size, they can account for the effects neither of length scale parameters nor of any structural change at the micro-level induced by deformation.
(ii) In the method mentioned in (ii) one needs to postulate suitable balance equations from which deducing the evolution equations for *micro-descriptors*. No general axiomatic format seems presently available to supply valid criteria to select the appropriate form for these balance equations. For a more detailed discussion of this point we refer to [22], [23].
(iii) On the other hand the procedure proposed in [2] supplies a standard method for simplifying the *refined* description for $\mathscr{S}$ by means of the introduction of a micro-structured continuum, modelling it at a *macroscopic level*. In particular the aforementioned procedure allows for an easy determination of both

- the evolution equations for the macro displacement and the relevant micro descriptors introduced for describing the kinematics of the micro structured continuum characterizing the *coarse* behavior of $\mathscr{S}$,
- and the constitutive equations for the micro structured continuum in terms of the geometry of $\mathscr{S}$ at the micro-level and the constitutive equations of its micro-component.

The method used here is based on the deduction of evolution equations for both macro displacements and micro descriptors from the classical theory of Cauchy continua, assumed to be valid in the refined description at the micro level. The micro-displacement field for these continua is found — in principle — by using force balance law, when the geometry of the solid matrix, its interaction modalities with the field phase, and the constitutive equations for both solid and fluid phases are known.

Although their determination can be very difficult, in many instances it is possible to guess some features of the micro-displacements in order to obtain a simplified system of PDE governing the motion. In [2] this guessing procedure finds a formally correct frame of reference by means of the introduction of so called "micro-shape functions". Their properties are listed in the following Sections 2, 3 or in [1], [2] to which we refer. The suitably chosen micro-shape functions allow for the representation of micro-displacement as the sum of a macro-displacement and a micro-disturbance whose evolution is controlled by the micro descriptors and whose form is fixed by micro-shape functions. In this way it is possible to account for those features of the deformation of the micro-structure relevant for the evolution of macro displacement. Indeed the weak form of the force balance equation — which we assume to hold at the micro-level — once the aforementioned representation for micro-displacement is accepted, allows for the deduction of the evolution equations for macro displacements and micro descriptors.

This deduction — which makes possible the change from the refined to a coarse description of $\mathscr{S}$ — is based on

- the determination of an accuracy parameter $\varepsilon$ and of the micro-periodicity cells;
- the assumption that macro displacements and micro descriptors are slowly-varying fields. More precisely macro displacements and micro descriptors are assumed to be $\varepsilon$-macro functions, i.e. functions whose variation in every periodicity cell is $\varepsilon$-negligible;
- the possibility, in the macro description of $\mathscr{S}$, of neglecting — in the weak form of force balance — every quantity of order equal or greater than $\varepsilon$.

While the equations for macro displacements are PDE, those obtained for micro descriptors are ODE. Therefore — in the sense used by [16] — the micro descriptors are internal state variables.

In [2] also a criterion is given to establish whether the chosen micro shape functions are reasonably guessed: for a discussion of the reliability of the micro shape functions we use in this paper we refer to [1]–[3].

In the present paper we:

- formulate a new mathematical model for compacting fluid saturated grounds
- obtain, in the framework of the aforementioned model, the description of the genesis of the microearthquakes related to the bradyseismic phenomena in the Phlegrean Fields in South Italy.

## 2 Geometrical structure of the solid matrix and fluid inclusions

The considered body $\mathscr{B}$ is assumed to be constructed — in a reference configuration — by translating a typical cell $V$. Let

$$V := \left(-\frac{l_1}{2}, \frac{l_1}{2}\right) \times \left(-\frac{l_2}{2}, \frac{l_2}{2}\right) \times \left(-\frac{l_3}{2}, \frac{l_3}{2}\right); \quad l := \sqrt{\sum_{i=1}^{3} l_i^2}. \tag{1}$$

We choose $l$ as the micro-structure length scale parameter (cf. [2] and references there quoted).

In order to characterize the structure of a solid matrix with non connected fluid inclusions we consider the following partition of $V$:

$$V = V_F \cup V_S, \quad V_F \cap V_S = \emptyset$$

where

$$V_F \subseteq V \quad \text{such that} \quad \partial V_F \cap \partial V_S = \emptyset.$$

We assume that $V$, $V_S$ and $V_F$ are regular regions as defined by Truesdell [14]. $V_S$ is occupied by a solid material while $V_F$ by a fluid one. For an arbitrary point $\mathbf{x}$ in the physical 3-space $E$ we will use the following notation:

(i) $V(\mathbf{x}) := \mathbf{x} + V, \quad V_S(\mathbf{x}) := \mathbf{x} + V_S, \quad V_F(\mathbf{x}) := \mathbf{x} + V_F$.

Let $\Omega$ be the region occupied by the body $\mathscr{B}$ in the reference configuration. We assume that there exists a lattice $\Lambda$ of places in $E$ such that (where int $A$ means the topological interior of $A$)

(ii) $\bigl(\forall (\mathbf{x}_1, \mathbf{x}_2) \in \Lambda \times \Lambda\bigr)\, [\mathbf{x}_1 \neq \mathbf{x}_2 \rightarrow \text{int } V(\mathbf{x}_1) \cap \text{int } V(\mathbf{x}_2) = \emptyset]$;

$\Omega = \text{int} \bigcup_{x \in \Lambda} \bar{V}(\mathbf{x})$,    is a regular region.

Moreover

(iii) $\Omega_S := \text{int} \bigcup_{x \in \Lambda} \bar{V}_S(\mathbf{x})$,    $\Omega_F := \text{int} \bigcup_{x \in \Lambda} \bar{V}_F(\mathbf{x})$,    $\Omega^o := \{\mathbf{x} \in \Omega : V(\mathbf{x}) \subset \Omega\}$,

will represent, respectively, the placement of solid matrix, of fluid inclusions and of what is called the macro interior of $\Omega$.

(iv) The smallest characteristic length dimension $L$ of $\Omega$ satisfies

$$\frac{l}{L_\Omega} \ll 1.$$

A detailed (*refined* following [3]) dynamical description of the body $\mathscr{B}$ at a micro-level is obtained by introducing displacement fields whose variations in every cell $V(x)$ (whose diameter is $l$) can be relevant. Instead a coarse description, i.e., a description at a macro-level (in which only some overall "averaged" properties of the body are to be considered) will be obtained by introducing "macro descriptors" of displacement which are nearly constant inside every $V(x)$.

(v) For a generic function $f$ defined respectively in $V(x)$, $V_S(x)$ and $V_F(x)$ we denote the mean value as follows [3], [1]:

$$\langle f \rangle_\mathbf{x} := \frac{1}{|V|} \int_{V(\mathbf{x})} f\, dV, \quad \langle f \rangle_\mathbf{x}^S := \frac{1}{|V|} \int_{V_S(\mathbf{x})} f\, dV, \quad \langle f \rangle_\mathbf{x}^F := \frac{1}{|V|} \int_{V_F(\mathbf{x})} f\, dV.$$

If $f$ is a $V$-periodic function, then, $\forall \mathbf{x} \in E$ we obtain

$$\langle f \rangle_\mathbf{x} = \langle f \rangle, \quad \langle f \rangle_\mathbf{x}^S = \langle f \rangle^S, \quad \langle f \rangle_\mathbf{x}^F = \langle f \rangle^F$$

where $\langle f \rangle, \langle f \rangle^S, \langle f \rangle^F$ are constants.

(vi) Following [2] we call *micro-shape functions* a system of sufficiently regular $V$-periodic linear-independent functions $h_l^A(\cdot)$, $A = 1, \ldots, N$, representing a functional basis for a finite dimensional subspace of the space of continuous $V$-periodic functions such that

$$h_l^A(\mathbf{x}) \in O(l), \quad \nabla h_l^A(\mathbf{x}) \in O(1), \quad \langle h_l^A \rangle^S = \langle h_l^A \rangle^F = 0, \quad \mathbf{x} \in E, \tag{2}$$

where the index $l$ displays the dependence of the micro shape functions on the length scale parameter; $O(l)$ represents a set of functions of $l$ which are infinitesimal of equal or higher order than $l$, and $O(1)$ is a set of functions of $l$ whose maximum value remains finite when $l \rightarrow 0$.

(vii) For some $\varepsilon > 0$, we call a real-valued function $F$ defined on $\Omega$ an $\varepsilon$-macro-function if

$$(\forall \mathbf{x} \in \Omega^o)\, \bigl(\forall z \in V(\mathbf{x})\bigr)\, F(z) \cong F(\mathbf{x}) \quad \text{where } a \cong b \text{ means } \|a - b\| \leq \varepsilon. \tag{3}$$

For each $F$ the parameter $\varepsilon$ will be considered as an error implied by the choice of giving up the microscopically accurate description of the variation in $V(\mathbf{x})$ of $F$. Although in general different $F$ have different $\varepsilon$ (i.e. $\varepsilon = \varepsilon_F$), in the following we assume that one and the same $\varepsilon$ can be chosen for all macro-functions necessary to the description of the system.

## 3 Evolution equations for macro-displacements and macro-descriptors

We assume that $\mathscr{B}$ consists of

- a linear elastic solid matrix
- saturated inclusions of a Stokesian incompressible viscous fluid [15].

Therefore, at the micro-level we assume that the following constitutive equations hold: (we use the conventions $\mathbf{C}.\mathbf{v} := C_{ijkl}v_l$, $\mathbf{C}:\mathbf{w} := C_{ijkl}w_{kl}$ with summation over doubly repeated indices)

$$\mathbf{t}_S = \mathbf{C}:\mathbf{e} + \mathbf{D}:\dot{\mathbf{e}}; \quad \mathbf{e} = \mathrm{Sym}\big(\mathrm{grad}\,(\mathbf{u})\big); \quad \text{for the solid,} \tag{4}$$

$$\mathbf{t}_F = p\mathbf{I} + 2\tilde{\mu}_F\left(\mathrm{Sym}\big(\mathrm{grad}\,(\mathbf{v})\big) - \frac{1}{3}\mathrm{div}\,(\mathbf{v})\,I\right) \quad \text{for the fluid,} \tag{5}$$

where $\mathbf{u}$ is the micro-displacement vector field defined in $\Omega_S$ for every substantial point belonging to the matrix; $\mathbf{v}$ is the micro-velocity field in the fluid inclusions. Since $\|\nabla\mathbf{u}\| \ll 1$ we may set $\dot{\mathbf{e}} = \mathrm{sym}\,(\mathrm{grad}\,(\mathbf{v}))$; $\mathbf{C}$ is the linear elasticity fourth order tensor; $\mathbf{D}$ is the Navier-Stokes linear viscosity tensor; $p$, $\mathbf{t}_S$, $\mathbf{t}_F$, $\tilde{\mu}_F$ are the hydrostatic pressure field, the Cauchy tensor for the solid and the fluid media, the Stokes viscosity in the fluid inclusions, and $\mathbf{I}$ is the identity tensor, respectively.

Let $\varrho_S$, $\varrho_F$ represent mass density fields of the solid and fluid components, and let $\mathbf{b}$ be the external body force.

We assume the following weak-form of force balance equations (see for instance [16], [17]):

$$\int_{\Omega_S} \mathbf{t}_S : \delta\mathbf{e}\,dV + \int_{\Omega_F} \mathbf{t}_F : \delta\mathbf{e}\,dV + \frac{d}{dt}\left(\int_{\Omega_S}\varrho_S\dot{\mathbf{u}}\cdot\delta\mathbf{u}\,dV + \int_{\Omega_F}\varrho_F\dot{\mathbf{u}}\cdot\delta\mathbf{u}\,dV\right)$$

$$= \int_{\Omega_S}\varrho_S\mathbf{b}\cdot\delta\mathbf{u}\,dV + \int_{\Omega_F}\varrho_F\mathbf{b}\cdot\delta\mathbf{u}\,dV \tag{6}$$

for every $\delta\mathbf{u}$ continuous in $\Omega$ and of class $C^1$ in $\mathrm{int}\,\Omega_S \cup \mathrm{int}\,\Omega_F$, and such that $\delta\mathbf{u}|_{\partial\Omega} = 0$, $\delta\mathbf{e} = \mathrm{Sym}\,(\mathrm{grad}\,(\delta\mathbf{u}))$.

The dynamical problem for $\mathscr{B}$ is given by Eqs. (4), (5), and (6). It describes completely the micro-behavior of the considered continuum. No qualitative study of the resulting PDE has been performed. In particular up to now it has not been possible to describe the influence of the geometry of the fluid inclusions on the average displacement of the cells constituting the solid matrix. Only numerical solutions could be determined for particular geometrical shapes of the solid matrix.

In this paper we try the aforementioned qualitative study by refraining from a detailed description of the micro-displacement field inside the typical cell $V(x)$. Indeed, we guess a representation for the micro-displacement field in terms of a macro-displacement field and some disturbances obtained by means of some micro descriptors and some micro shape functions. The micro shape functions account for the geometry of the inclusions and for some aspects of the mechanical behavior of the solid matrix, while the micro-descriptors account for those aspects of the micro-displacements which are expected to influence the macro-displacements. Formally this can be done by introducing the following hypotheses (for more details we refer to [1]–[3], [12]):

(j) The micro-displacement field $u$ is represented as (summation over $A$ holds):

$$\mathbf{u}(\mathbf{x}, t) = \mathbf{U}(\mathbf{x}, t) + h_l^A(\mathbf{x})\, \mathbf{Q}^A(\mathbf{x}, t), \quad \mathbf{x} \in \Omega, \quad t \in (t_0, t_f), \tag{7}$$

where $h_l^A(\cdot)$, $(A = 1, ..., N)$ are the micro shape functions describing the kinematics of the system at the microscopic level. They are to be chosen a priori in order to account for those features of the solutions at the micro-level which are relevant at the macro-level in the considered phenomena. The more suitable the guess represented by the micro shape functions is the more accurate will be the qualitative analysis they allow. For instance, in [1] it is shown that for empty inclusions the most appropriate micro shape functions differ from those which have to be chosen in the case of filled inclusions. More examples are available in [3], [12], [2].

**Remark 1.** *The functions U and $\mathbf{Q}^A$ are sufficiently regular macro-functions which will be treated as new kinematical, independent variables. They will be called, respectively, the macro-displacement and the micro-descriptors of micro-displacement fields.*

(jj) In modelling the macro-behavior of the system we dispence with the description of some of the aspects that can be considered *micro*. Consequently, we neglect all terms appearing in the equations that are of the same (or higher) order of the accuracy parameter $\varepsilon$ that characterizes the precision to which we can measure the macro-functions $U(\cdot, t)$, $\mathbf{Q}^A(\cdot, t)$ and all their derivatives. In the sequel this kind of approximation will be denoted by $\cong$. It is better to underline at this point that in this approach the micro-structure length scale parameter $l$ is a known physical constant (independent of $\varepsilon$) and will not be neglected.

**Remark 2.** *Because of the definition of micro shape functions given in Section 2 we obtain*

$$\langle \mathbf{u} \rangle_\mathbf{x}^S = \langle \mathbf{u} \rangle_\mathbf{x}^F = \mathbf{U}(\mathbf{x}, t), \quad \forall \mathbf{x} \in \Omega^o, \quad \forall t \in (t_0, t_f);$$

*therefore the macro-motion of the matrix is the same as the macro-motion of the fluid inclusions. The hypotheses formulated for the micro shape functions are therefore suitable for describing the behavior of a solid matrix with non-connected fluid inclusions. This fact allows to use a Lagrangian description also for the fluid component of the considered system.*

In this way we see that at micro-level the motion, described by $\mathbf{u}$, can be regarded as a superimposition of "micro-disturbances" ($h_l^A \mathbf{Q}^A$) over the "macro-motion" $\mathbf{U}$. Thus the $\mathbf{Q}^A(\cdot, t)$ are those macro-quantities describing those disturbances at the micro-level which are relevant at the macro-level. Choosing different micro shape functions allows for the study of different aspects of the considered process, whose more detailed description can also be obtained by increasing the number $N$ of micro shape functions.

(jjj) In what follows we will assume, coherently with (7),

$$\delta \mathbf{u} = \delta \mathbf{U} + h_l^A \delta \mathbf{Q}^A \tag{8}$$

$$\mathbf{E} = \text{Sym}\,(\nabla \mathbf{U})$$

$$\mathbf{e} = \mathbf{E} + \text{Sym}\,(\nabla(h_l^A \mathbf{Q}^A)) \cong \mathbf{E} + \text{Sym}\,(\nabla h_l^A \otimes \mathbf{Q}^A)$$

$$\delta \mathbf{e} \cong \delta \mathbf{E} + \text{Sym}\,(\nabla h_l^A \otimes \delta \mathbf{Q}^A)$$

where $\delta \mathbf{U}$ and $\delta \mathbf{Q}^A$ are $\varepsilon$-macro-functions.

Using (j), (jj), and (jjj) we obtain, by manipulating each term in Eq. (6), the weak form of the evolution equations for $\mathbf{U}$ and $\mathbf{Q}^A$. To this end observe that (for a detailed derivation of these expressions we refer to Eq. (61) in the Appendix)

$$\int_{\Omega_F} (\mathbf{t}_F - p\mathbf{I}) : \delta\mathbf{e}\, dV + \int_{\Omega_S} \mathbf{t}_S : \delta\mathbf{e}\, dV \cong \int_{\Omega} (\mathbf{S} : \delta\mathbf{E} + \mathbf{H}^A \cdot \delta\mathbf{Q}^A)\, dV \tag{9}$$

where $\forall t \in (t_0, t_f)$ and $\forall \mathbf{x} \in \Omega^o$ we define the following system of macro-internal forces:

$$\mathbf{S}(\mathbf{x}, t) := \langle \mathbf{t}_S \rangle_\mathbf{x}^S + \langle \mathbf{t}_F - p\mathbf{I} \rangle_\mathbf{x}^F, \quad \mathbf{H}^A(\mathbf{x}, t) := \langle \mathbf{s}.\nabla h_l^A \rangle_\mathbf{x}^S. \tag{10}$$

Because of the last equation we obtain the following macro constitutive equations:

$$\mathbf{S} \cong \langle \mathbf{C} \rangle : \mathbf{E} + \langle \mathbf{C}.\nabla h_l^A \rangle .\mathbf{Q}^A + \langle \mathbf{D} \rangle : \dot{\mathbf{E}} + \langle \mathbf{D}.\nabla h_l^A \rangle .\dot{\mathbf{Q}}^A \tag{11}$$

$$\mathbf{H}^A \cong \langle \mathbf{C}.\nabla h_l^A \rangle : \mathbf{E} + \langle \mathbf{C} : (\nabla h_l^A \otimes \nabla h_l^B) \rangle .\mathbf{Q}^B + \langle \mathbf{D}.\nabla h_l^A \rangle : \dot{\mathbf{E}} + \langle \mathbf{D} : (\nabla h_l^A \otimes \nabla h_l^B) \rangle .\dot{\mathbf{Q}}^B. \tag{12}$$

**Remark 3.** *The last two equations imply that both $\mathbf{S}$ and $\mathbf{H}^A$ are macro-functions in $\Omega$. Moreover, we underline that the tensor S because of its definition accounts also for the dissipative component of the micro stress tensor in the fluid inclusions.*

In the same way we find (see Eqs. (64) and (65) in the Appendix):

$$\int_{\Omega_F} \varrho \dot{\mathbf{u}} \cdot \delta\mathbf{u}\, dV + \int_{\Omega_S} \varrho \dot{\mathbf{u}} \cdot \delta\mathbf{u}\, dV \cong \int_{\Omega} (\langle \varrho \rangle \dot{\mathbf{U}} + \langle \varrho h_l^A \rangle \dot{\mathbf{Q}}^A) \cdot \delta\mathbf{U}\, dV$$

$$+ \int_{\Omega} (\langle \varrho h_l^A \rangle \dot{\mathbf{U}} + \langle \varrho h_l^A h_l^B \rangle \dot{\mathbf{Q}}^B) \cdot \delta\mathbf{Q}^A\, dV \tag{13}$$

$$\int_{\Omega_F} \varrho \mathbf{b} \cdot \delta\mathbf{u}\, dV + \int_{\Omega_S} \varrho \mathbf{b} \cdot \delta\mathbf{u}\, dV \cong \int_{\Omega} [\langle \varrho \mathbf{b} \rangle \cdot \delta\mathbf{U} + \langle \varrho \mathbf{b} h_l^A \rangle \cdot \delta\mathbf{Q}^A]\, dV, \tag{14}$$

$$\int_{\Omega_F} p\, \mathrm{div}\,(\delta\mathbf{u})\, dV \cong \int_{\Omega} (-\nabla P \cdot \delta\mathbf{U} + \mathbf{P}^A \cdot \delta\mathbf{Q}^A)\, dV \tag{15}$$

where

$$P = \langle p \rangle_\mathbf{x}^F, \quad \mathbf{P}^A = \langle p \nabla h_l^A \rangle_\mathbf{x}^F, \quad \forall \mathbf{x} \in \Omega^o. \tag{16}$$

Therefore we obtain the following weak form for the PDE and ODE governing, respectively, the evolution of $\mathbf{U}$ and $\mathbf{Q}^A$:

$$\int_{\Omega} (\mathbf{S} : \delta\mathbf{E} + \mathbf{H}^A \cdot \delta\mathbf{Q}^A)\, dV + \int_{\Omega} (\langle \varrho \rangle \ddot{\mathbf{U}} + \langle \varrho h_l^A \rangle \ddot{\mathbf{Q}}^A) \cdot \delta\mathbf{U}\, dV$$

$$+ \int_{\Omega} (\langle \varrho h_l^A \rangle \ddot{\mathbf{U}} + \langle \varrho h_l^A h_l^B \rangle \ddot{\mathbf{Q}}^B) \cdot \delta\mathbf{Q}^A\, dV$$

$$= \int_{\Omega} [(\langle \varrho \mathbf{b} \rangle \cdot \delta\mathbf{U} + \langle \varrho \mathbf{b} h_l^A \rangle \cdot \delta\mathbf{Q}^A) + (\nabla P \cdot \delta\mathbf{U} + \mathbf{P}^A \cdot \delta\mathbf{Q}^A)]\, dV \tag{17}$$

$\forall \delta\mathbf{U}, \delta\mathbf{Q}^A, \delta\mathbf{U}|_{\partial\Omega} = 0, \quad \delta\mathbf{Q}^A|_{\partial\Omega} = 0.$

The local differential form of Eq. (17) reads

$$\mathrm{div}\,\mathbf{S} - \langle \varrho \rangle \ddot{\mathbf{U}} - \langle \varrho h_l^A \rangle \ddot{\mathbf{Q}}^A + \nabla P + \langle \varrho \mathbf{b} \rangle = 0 \tag{18}$$

$$\langle \varrho h_l^A h_l^B \rangle \ddot{\mathbf{Q}}^B + \langle \varrho h_l^A \rangle \ddot{\mathbf{U}} + \mathbf{H}^A + \mathbf{P}^A - \langle \varrho \mathbf{b} h_l^A \rangle = 0. \tag{19}$$

Equations (18), (19) together with (11), (12) and (16) represent the system of equations for the fields $\mathbf{U}$ and $\mathbf{Q}^A$, provided that the terms in which the fluid pressure in an inclusions appears are known. In the following we will limit ourselves to consider incompressible fluids. This implies that

$$\langle \mathrm{div}\,(\mathbf{u}) \rangle = 0 \Rightarrow \langle 1 \rangle^F \,\mathrm{div}\,\mathbf{U} + \langle \nabla h_l{}^A \rangle^F \cdot \mathbf{Q}^A = 0. \tag{20}$$

**Remark 4.** *The coefficients of the system (18), (19) and (20) are constant. This is a consequence of the hypothesis of V-periodicity we have made on the solid matrix. Note, also, that some of the aforementioned coefficients depend on l because of the terms*

$$\langle \varrho h_l{}^A h_l{}^B \rangle \sim l^2, \quad \langle \varrho h_l{}^A \rangle \sim l.$$

**Remark 5.** *Equation (19) represents a set of ODE. Therefore there is no boundary condition to assign for the $\mathbf{Q}^A$: in the sense of [16] they must be regarded as internal state variables.*

**Remark 6.** *Simple inspection of Eqs. (12) and (19) shows that it is not possible to use them to express S — as given by (11) — as functions of the quantities $\mathbf{E}$ and $\dot{\mathbf{E}}$ only. Indeed if the length scale parameter l is not negligible, (19) (after substitution of (12)) becomes an ODE of the second order for the variables $\mathbf{Q}^A$. If l is negligible Eq. (19) reduces to an ODE of the first order for the $\mathbf{Q}^A$'s. In both cases the possibility to represent S — by means of suitable Effective Moduli (as in the mogenized theories) — as a function of macro-strain and its time derivative is lost.*

## 4 A compaction problem: the effect of fluid inclusions on vibrations induced by ground settling

Let us apply the model to the case in which the solid matrix undergoes a compaction process. In particular we wish to study the effect of the geometry of the microscopic fluid inclusions on the micro-displacement waves induced by ground settling. To our knowledge this problem cannot be handled with other methods. We assume that $\mathbf{C}$ and $\mathbf{D}$ act on $\mathbf{E}$ and $\dot{\mathbf{E}}$ as follows:

$$\mathbf{C}:\mathbf{E} = \begin{cases} 0, & \text{if } \mathbf{x} \in \Omega_F \\ 2\mu_S \mathbf{E} + \lambda_S \,\mathrm{Tr}\,(\mathbf{E})\,\mathbf{I}, & \text{if } \mathbf{x} \in \Omega_S \end{cases} \tag{21}$$

$$\mathbf{D}:\dot{\mathbf{E}} = \begin{cases} 2\tilde{\mu}_F \left( \dot{\mathbf{E}} - \dfrac{1}{3}\,\mathrm{Tr}\,(\dot{\mathbf{E}})\,\mathbf{I} \right), & \text{if } \mathbf{x} \in \Omega_F \\ 2\tilde{\mu}_S \left( \dot{\mathbf{E}} - \dfrac{1}{3}\,\mathrm{Tr}\,(\dot{\mathbf{E}})\,\mathbf{I} \right), & \text{if } \mathbf{x} \in \Omega_S. \end{cases}$$

We will distinguish two regimes, separate in time: the first ($t \in [0, t_0]$) in which the solid compacts in the $x_1$ direction. Because of compression some micro-motions arise which are able to influence the behavior of the body at the macro level also. We study these effects in an interval following the compaction: $t \in [t_0, t_f]$, (eventually ($t_f \to \infty$)), analyzing the influence of the micro structure size on the macro-behavior and compare our results with those obtained with the homogenized asymptotic model.

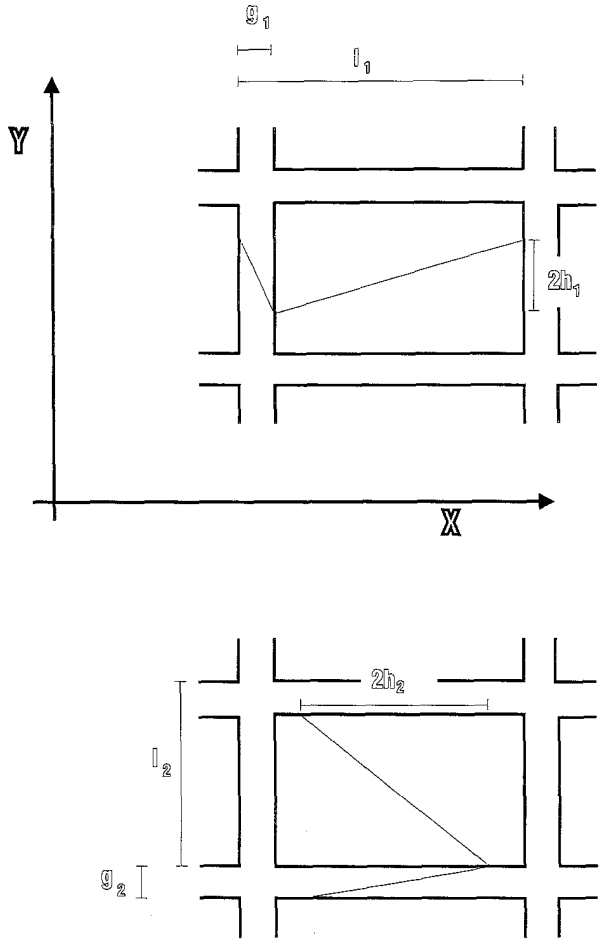

Fig. 1. Shape and size of a typical cell, plots of used micro-shape-functions

The initial conditions for $t = t_0$ are determined as the final conditions of the compaction regime so that we must necessarily analyze the compaction process in some detail. In general in order to model these processes (see [18]–[20]) PDE in three dimensional domains need to be solved. Here we use a semi-inverse method for solving the system of PDE (for MD) coupled with the ODE governing the evolution of micro-descriptors. We use the following micro-shape functions (see Fig. 1):

$$h^i(x_i) = \begin{cases} h_i - 2\dfrac{h_i}{g_i} x_i, & \text{if } 0 < x_i < g_i, \\ -h_i + 2h_i \dfrac{x_i - g_i}{l_i - g_i}, & \text{if } g_i < x_i < l_i, \end{cases} \quad i = 1, 2, 3. \tag{22}$$

This choice has been fully justified from the mathematical point of view in [1]–[3], [12]. Here we want to stress the physical meaning: it states that the unit $V$-cell and its fluid inclusion $V_F$ undergo deformations mapping them into parallelepipeds only. This assumption is well-grounded only if a solid matrix is in contact with an incompressible or nearly incompressible fluid inclusion. For instance if we apply our modelling procedure to a solid with voids then we need as micro shape functions higher order polynomials in the x variables.

*4.1 The initial phase: ground settling*

We use the following boundary conditions (BC) for the macro displacement:

$$U_1(0, x_2, x_3, t) = 0, \qquad U_1(L_1, x_2, x_3, t) = -\eta L_1 \frac{t}{t_0},$$

$$U_1(x_1, 0, x_3, t) = -\eta x_1 \frac{t}{t_0}, \quad U_1(x_1, L_2, x_3, t) = -\eta x_1 \frac{t}{t_0}, \tag{23}$$

$$U_1(x_1, x_2, 0, t) = -\eta x_1 \frac{t}{t_0}, \quad U_1(x_1, x_2, L_3, t) = -\eta x_1 \frac{t}{t_0},$$

$$U_2 = U_3 = 0 \quad \text{on} \quad \partial\Omega$$

where $\eta$ is a measure of the amount of the ground settling. Indeed the total variation of the ground level $\delta$ at $x_3 = 0$ is given by $\delta = \eta L_1$.

To these boundary conditions we must add the following initial conditions (IC):

$$U_1(x_1, x_2, x_3, 0) = 0, \qquad \dot{U}_1(x_1, x_2, x_3, 0) = -\frac{\delta_0}{t_0} \frac{x_1}{L_1},$$

$$U_2(x_1, x_2, x_3, 0) = 0, \qquad \dot{U}_2(x_1, x_2, x_3, 0) = 0,$$

$$\dot{U}_2(x_1, x_2, x_3, 0) = 0, \qquad \dot{U}_3(x_1, x_2, x_3, 0) = 0, \tag{24}$$

$$Q_i^A(x_1, x_2, x_3, 0) = 0, \quad (A = 1, 2, 3; i = 1, 2, 3),$$

$$\dot{Q}_i^A(x_1, x_2, x_3, 0) = 0, \quad (A = 1, 2, 3; i = 1, 2, 3).$$

It is very easy to verify that

$$U_1 = -\delta_0 \frac{x_1}{L_1} \frac{t}{t_0}, \quad U_2 = U_3 = 0, \quad P = P(t) \tag{25}$$

is a solution of

$$\operatorname{div} \mathbf{S} - \langle \varrho \rangle \ddot{\mathbf{U}} + \nabla P = 0 \tag{26}$$

satisfying BC and IC.

In this way Eqs. (18), (19) and (20) simplify and can be written

$$\langle \varrho(h^1)^2 \rangle \ddot{Q}_1^1 + H_1^1 + P_1 = 0, \tag{27}$$

$$\langle \varrho(h^2)^2 \rangle \ddot{Q}_2^2 + H_2^2 + P_2 = 0, \tag{28}$$

$$\langle \varrho(h^3)^2 \rangle \ddot{Q}_3^3 + H_3^3 + P_3 = 0, \tag{29}$$

$$e_1 Q_1^1 + e_2 Q_2^2 + e_3 Q_3^3 = \langle 1 \rangle^F \eta \frac{t}{t_0}, \tag{30}$$

with

$$P_i = P \langle \partial_i h^i \rangle_F, \quad (i = 1, 2, 3), \tag{31}$$

$$H_1^1 = f_1 \partial_1 U_1 + g_1 \partial_1 \dot{U}_1 + C_{11} Q_1^1 + C_{12} Q_2^2 + C_{13} Q_3^3 + D_{11} \dot{Q}_1^1 + D_{12} \dot{Q}_2^2 + D_{13} \dot{Q}_3^3, \tag{32}$$

A micro-structured continuum 175$$H_2{}^2 = f_2\,\partial_1 U_1 + g_2\,\partial_1 \dot{U}_1 + C_{22}Q_2{}^2 + C_{21}Q_1{}^1 + C_{23}Q_3{}^3 + D_{22}\dot{Q}_2{}^2 + D_{21}\dot{Q}_1{}^1 + D_{23}\dot{Q}_3{}^3, \tag{33}$$

$$H_3{}^3 = f_3\,\partial_1 U_1 + g_3\,\partial_1 \dot{U}_1 + C_{33}Q_3{}^3 + C_{31}Q_1{}^1 + C_{32}Q_2{}^2 + D_{33}\dot{Q}_3{}^3 + D_{31}\dot{Q}_1{}^1 + D_{32}\dot{Q}_2{}^2, \tag{34}$$

where

$$\langle \varrho(h^i)^2 \rangle = \frac{(h^i)^2\, l^2}{3}(g_i{}^3 \varrho_S + (1-\vartheta_i)^3 \varrho_F), \quad (i=1,2,3), \tag{35}$$

$$e_i = \langle \partial_i h^i \rangle_F, \quad (i=1,2,3),$$

$$f_1 = \langle (\lambda_S + 2\mu_S)(\partial_1 h^1) \rangle_S,$$

$$g_1 = \frac{4}{3}(\langle \tilde{\mu}_S(\partial_1 h^1)\rangle_S + \langle \tilde{\mu}_F(\partial_1 h^1)\rangle_F),$$

$$f_2 = \langle \lambda_S(\partial_2 h^2)\rangle_S,$$

$$g_2 = -\frac{2}{3}(\langle \tilde{\mu}_S(\partial_2 h^2)\rangle_S + \langle \tilde{\mu}_F(\partial_2 h^2)\rangle_F),$$

$$f_3 = \langle \lambda_S \rangle (\partial_3 h^3)_S,$$

$$g_3 = -\frac{2}{3}(\langle \tilde{\mu}_S(\partial_3 h^3)\rangle_S + \langle \tilde{\mu}_F(\partial_3 h^3)\rangle_F),$$

$$C_{ii} = \langle (\lambda_S + 2\mu_S)(\partial_i h^i)^2 \rangle_S, \quad (i=1,2,3), \tag{36}$$

$$C_{ij} = \langle \lambda_S (\partial_i h^i)(\partial_j h^j)\rangle_S, \quad (i,j=1,2,3;\, i\neq j),$$

$$D_{ii} = \frac{4}{3}[\langle \tilde{\mu}_S(\partial_i h^i)^2\rangle_S + \langle \tilde{\mu}_F(\partial_i h^i)^2\rangle_F] \quad (i=1,2,3),$$

$$D_{ij} = -\frac{2}{3}[\langle \tilde{\mu}_S(\partial_i h^i)(\partial_j h^j)\rangle_S + \langle \tilde{\mu}_F(\partial_i h^i)(\partial_j h^j)\rangle_F], \quad (i,j=1,2,3;\, i\neq j).$$

To give an example of the possible application of our model we limit ourselves to the case in which

$$l_1 = l_2 = l_3 = l_0 = \frac{l}{\sqrt{3}} \tag{37}$$

$$h_1 = h_2 = h_3 = h l_0$$

$$g_1 = g_2 = g_3 = g l_0$$

corresponding to a cubic solid matrix with cubic inclusions. In this case we obtain the same equations along the $x_2$ and $x_3$ directions, so that Eqs. (27)–(30) reduce to a system of three differential equations ($Q_2{}^2 = Q_3{}^3 = Q_0$ and $Q_1{}^1 = Q_1$):

$$\langle \varrho(h^1)^2\rangle \ddot{Q}_1 + H_1{}^1 + P_1 = 0 \tag{38}$$

$$\langle \varrho(h^2)^2\rangle \ddot{Q}_0 + H_2{}^2 + P_2 = 0 \tag{39}$$

$$e_1 Q_1 + (e_2 + e_3) Q_0 = \langle 1 \rangle^F \eta \frac{t}{t_0}. \tag{40}$$

Calculating $Q_1$ and $P$ from Eqs. (40) and (38) and substituting in (33) and (39) we finally obtain

$$\ddot{Q}_0 + \alpha_0 \dot{Q}_0 + \beta_0 Q_0 = \gamma_0 \, \partial_1 U_1 + \gamma_1 \, \partial_1 \dot{U}_1 \tag{41}$$

with

$$\alpha_0 = 24 \frac{1}{l_0^2} \left[ \frac{g\tilde{\mu}_S + (1-g)\tilde{\mu}_F}{g^3 \varrho_S + (1-g)^3 \varrho_F} \right], \tag{42}$$

$$\beta_0 = 16 \frac{1}{l_0^2} \left[ \frac{g\mu_S}{g^3 \varrho_S + (1-g)^3 \varrho_F} \right], \tag{43}$$

$$\gamma_0 = -8 \frac{1}{h l_0^2} \left[ \frac{g^2 \mu_S}{g^3 \varrho_S + (1-g)^3 \varrho_F} \right], \tag{44}$$

$$\gamma_1 = \frac{4}{3h l_0^2} \left[ \frac{-g^2 \tilde{\mu}_S + (1-g)^2 \tilde{\mu}_F}{g^3 \varrho_S + (1-g)^3 \varrho_F} \right]. \tag{45}$$

The solution of Eq. (41) is (a similar expression for $Q_1$ is obtained using (40))

$$Q_0 = -\frac{\eta}{\beta_0 t_0}\left[\gamma_0\left(t - \frac{\alpha_0}{\beta_0}\right) + \gamma_1\right] + A_1 e^{(-\alpha_0 - \sqrt{\alpha_0^2 - 4\beta_0})\,(t/2)} + A_2 e^{(-\alpha_0 + \sqrt{\alpha_0^2 - 4\beta_0})\,(t/2)} \tag{46}$$

with

$$A_1 = \eta \frac{\alpha_0^2 \gamma_0 - 2\beta_0 \gamma_0 - \alpha_0 \beta_0 \gamma_1 - \sqrt{\alpha_0^2 - 4\beta_0}(\gamma_0 \alpha_0 - \beta_0 \gamma_1)}{2\beta_0^2 t_0 \sqrt{\alpha_0^2 - 4\beta_0}},$$

$$A_2 = \eta \frac{-2\beta_0 \gamma_0 + (\alpha_0 + \sqrt{\alpha_0^2 - 4\beta_0})(\gamma_0 \alpha_0 - \beta_0 \gamma_1)}{2\beta_0^2 t_0 \sqrt{\alpha_0^2 - 4\beta_0}}.$$

The qualitative behavior of $Q_0$ and $Q_1$ is determined by the sign of $\alpha_0$, $\beta_0$ and $\alpha_0^2 - 4\beta_0$. We see that both $\alpha_0 > 0$ and $\beta_0 > 0$. It remains to study the sign of $\alpha_0^2 - 4\beta_0$. If

$$l^2 < \frac{9(g\tilde{\mu}_S + (1-g)\tilde{\mu}_F)^2}{g\mu_S(g^3 \varrho_S + (1-g)^3 \varrho_F)}, \tag{47}$$

then

$$\alpha_0^2 - 4\beta_0 > 0; \tag{48}$$

so the exponential term in Eq. (46) is real and $Q_0(t)$ exponentially decreasing. Otherwise if (47) does not hold, $Q_0(t)$ will undergo damped oscillations.

### 4.2 The subsequent phase: micro-vibrations

We assume that in the interval $[t_0, t_f]$ the IC for the macro displacement and the microdescriptors are given by their corresponding final value in the settling phase. Therefore, when the settling process is ended, we have the following boundary conditions:

$$U_1(0, x_2, x_3, t) = 0, \qquad U_1(L_1, x_2, x_3, t) = -\delta,$$

$$U_1(x_1, 0, x_3, t) = -\eta x_1, \qquad U_1(x_1, L_2, x_3, t) = -\eta x_1,$$

$$U_1(x_1, x_2, 0, t) = -\eta x_1, \qquad U_1(x_1, x_2, L_3, t) = -\eta x_1, \qquad (49)$$

$$U_2 = U_3 = 0 \quad \text{on} \quad \partial\Omega,$$

together with the initial conditions

$$U_1(x_1, x_2, x_3, t_0) = -\eta x_1, \qquad \dot{U}_1(x_1, x_2, x_3, t_0) = 0,$$

$$U_2(x_1, x_2, x_3, t_0) = 0, \qquad \dot{U}_2(x_1, x_2, x_3, t_0) = 0,$$

$$U_3(x_1, x_2, x_3, t_0) = -0, \qquad \dot{U}_3(x_1, x_2, x_3, t_0) = 0, \qquad (50)$$

$$Q_i^A(x_1, x_2, x_3, t_0) = \bar{Q}_i^A, \qquad (A = 1, 2, 3; i = 1, 2, 3),$$

$$\dot{Q}_i^A(x_1, x_2, x_3, t_0) = \hat{Q}_i^A, \qquad (A = 1, 2, 3; i = 1, 2, 3),$$

where $\bar{Q}_i^A$ and $\hat{Q}_i^A$ are, respectively, the values of $Q_i^A$ and $\dot{Q}_i^A$ at the time $t = t_0$ as obtained in the previous Section. The solution of Eqs. (38), (40) is

$$Q_0(t) = -\eta \frac{\gamma_0}{\beta_0} + \frac{\eta}{2\beta_0 t_0 \sqrt{\alpha_0^2 - 4\beta_0}} \left[ B_1 e^{(-\alpha_0 - \sqrt{\alpha_0^2 - 4\beta_0}) \frac{t-t_0}{2}} + B_2 e^{(-\alpha_0 + \sqrt{\alpha_0^2 - 4\beta_0}) \frac{t-t_0}{2}} \right]$$

$$+ A_1 e^{(-\alpha_0 - \sqrt{\alpha_0^2 - 4\beta_0})(t/2)} + A_2 e^{(-\alpha_0 + \sqrt{\alpha_0^2 - 4\beta_0})(t/2)}, \qquad (51)$$

where

$$B_1 = 2\gamma_0 - \left( \frac{\alpha_0 \gamma_0}{\beta_0} - \gamma_1 \right) \left( -\alpha_0 + \sqrt{\alpha_0^2 - 4\beta_0} \right),$$

$$B_2 = -2\gamma_0 + \left( \frac{\alpha_0 \gamma_0}{\beta_0} - \gamma_1 \right) \left( -\alpha_0 - \sqrt{\alpha_0^2 - 4\beta_0} \right).$$

**Remark 1.** $Q_1(t)$ is obtained by substituting Eq. (51) in Eq. (40).

**Remark 2.** Observe that $\lim_{t \to \infty} Q_0(t) = -\eta \frac{\gamma_0}{\beta_0} \neq Q_0(t_0)$. This means that, at the micro level, after the first phase of ground settling a second settling phase — charaterized by micro vibrations — occurs.

**Remark 3.** If $\alpha_0^2 - 4\beta_0 < 0$ the solid matrix undergoes — in general — damped oscillations at the micro-level, along all three spatial directions. Therefore — in the framework of our model — we can prove that — under some particular geometrical and mechanical conditions, of the type represented by Eq. (47) with the reversed inequality sign — some damped displacement waves arise in the solid matrix — filled with a viscous fluid — as a consequence of a compaction process.

**Remark 4.** The pressure $P(t)$ can be obtained by one of the two Eqs. (38), (39), once the expressions of $Q_0(t)$ and $Q_1(t)$ are obtained. $P(t)$ undergoes damped oscillations if $Q_0(t)$ and $Q_1(t)$ do.

## 5 Homogenized asymptotic model

In this case the equations are readily obtained from (18) by letting $l \to 0$ in (19). Equations (38), (39) and (40) then become

$$C_{11}Q_1 + C_{10}Q_0 + D_{11}\dot{Q}_1 + D_{10}\dot{Q}_0 + C_1 P^1 = -f_1 \, \partial_1 U_1 - g_1 \, \partial_1 \dot{U}_1, \tag{52}$$

$$C_{01}Q_1 + C_{00}Q_0 + D_{01}\dot{Q}_1 + D_{00}\dot{Q}_0 + C_0 P^2 = -f_2 \, \partial_1 U_1 - g_2 \, \partial_1 \dot{U}_1, \tag{53}$$

$$Q_1{}^1 = -e_1 \, \partial_1 U_1 - (e_2 + e_3) Q_2{}^2 \tag{54}$$

and with the same calculations as in the previous cases we obtain, in the phase of ground settling,

$$\tilde{\alpha}_0 \dot{Q}_0 + \tilde{\beta}_0 Q_0 = \tilde{\gamma}_0 \, \partial_1 U_1 + \tilde{\gamma}_1 \, \partial_1 \dot{U}_1, \tag{55}$$

where $\tilde{\alpha}_0 = \alpha_0$, $\tilde{\beta}_0 = \beta_0$, $\tilde{\gamma}_0 = \gamma_0$, $\tilde{\gamma}_1 = \gamma_1$; therefore

$$Q_0(t) = -\eta \frac{\tilde{\gamma}_0}{\tilde{\beta}_0} \frac{t}{t_0} + A_3(1 - e^{-(\tilde{\beta}_0/\tilde{\alpha}_0)t}) \tag{56}$$

with $A_3 = \eta \left( \dfrac{\tilde{\alpha}_0 \tilde{\gamma}_0}{\tilde{\beta}_0{}^2 t_0} - \dfrac{\tilde{\gamma}_1}{\tilde{\beta}_0 t_0} \right)$. On the other hand, in the subsequent phase, we have

$$Q_0(t) = -\eta \frac{\tilde{\gamma}_0}{\tilde{\beta}_0} + A_3(e^{-(\tilde{\beta}_0/\tilde{\alpha}_0)t_0} - 1) \, e^{-(\tilde{\beta}_0/\tilde{\alpha}_0)/t}. \tag{57}$$

**Remark 5.** *Because of the dissipative terms appearing in the constitutive equations (4) and (5) and consequently in Eq. (19) also in the case of the homogenized asymptotic model the possibility of introducing effective elasticity moduli is lost.*

**Remark 6.** *Once the "pore-size" length scale parameter is neglected, i.e. once an homogenized asymptotic model is accepted, it is not possible to account for those cases in which damped oscillations arise after compaction.*

## 6 Conclusions

### 6.1 Comparison with the homogenized asymptotic models

We start by comparing the results of our refined "length scale parameter" model with the asymptotic model. Calculating the limit when $l \to 0$ we find from (42)–(45) that

$$-\alpha_0 - \sqrt{\alpha_0{}^2 - 4\beta_0} \to -\infty,$$

$$-\alpha_0 + \sqrt{\alpha_0{}^2 - 4\beta_0} \to -2\frac{\tilde{\beta}_0}{\tilde{\alpha}_0}, \tag{58}$$

$$A_2 \to A_3.$$

Thus, in this limit the solutions found in Section 4 turn into the asymptotic one found in Section 5.

We observe that within the homogenized asymptotic model that part of the general solution depending on $A_1$ cannot be obtained.

*6.2 Towards an application to the genesis of micro-earthquake induced by compaction*

We note that, because of the dependence of $\alpha_0$ and $\beta_0$ on $l$ and $g$, we can prove that there are two ranges in the set of geometric and material constants. Bodies characterized by constants belonging to the first set are such that all initial strain deformations are damped in an exponentially decreasing way in a short "transient" time interval. On the other hand media with material constants belonging to the second class experience a longer period of damped displacement oscillations.

The quoted results open interesting possibilities towards the description of some among the phenomena described in [4], [5] and [6]. These papers show how peculiar the seismic and volcanic activity is in the region of Phlegraean Fields in southern Italy. In particular in [5] micro-earthquakes are decribed which are typical of that area.

On the other hand in [4] — in a very intuitive and somehow tentative way — some ideas are presented about the genesis of those earthquakes and the whole bradyseismic[1] phenomenon. The ideas are:

(i) the vertical movement of the crust in the Phlegraean Fields could be caused by the pore pressure variation in the porous fluid saturated media filling their volcanic basin.
(ii) the micro-earthquakes accompanying both the positive (crustal uplift) and the negative (crustal settling) phases of the bradyseism could be a consequence of a sudden micro-strain release following by ground settling.

It seems to us that the results of the present paper allow us to perceive the formulation of a mathematical model suitable for the qualitative and quantitative substantiation of the idea mentioned in (ii). Indeed we are already able to prove that micro-displacement and *macro-pressure* damped waves can arise — under particular conditions — as a consequence of ground settling. On the other hand, some improvements of the present treatment need to be performed: in particular, a model capable to account for the phenomena occurring in solid matrices with *micro-connected* fluid inclusions need to be formulated. This generalization will be presented in a future paper.

We conclude by underlining that the ideas mentioned in (ii) can be found — in a more intuitive, but also more suggestive form — already in a very ancient tractate, by Lucretius:

*Nunc age quae ratio terrai motibus exstet percipe. et in primis terram fac ut esse rearis subter item ut supera ventosis undique plenam speluncis multosque et rupis deruptaque saxa multaque sub tergo terrai flumina tecta volvere vi fluctus summersaque saxa putandumst. undique enim similem esse sui res postulat ipsa. his igitur rebus subiunctis suppositisque terra superne tremit magnis concussa ruinis, subter ubi ingentis speluncas subruit aetas; quippe cadunt toti montes magnoque repente concussu late disserpunt inde tremores.*

<div style="text-align:right">Lucretius, De Rerum Natura Liber VI, 535–547</div>

We believe to be useful to the reader quoting here the English translation of this extract by Rouse and Ferguson Smith [24]:

*Now attend and learn what is the reason for earthquakes. And in the first place, be sure to consider the earth below as above to be everywhere full of windy caverns, bearing many lakes and many pools in her bosom with rocks and steep cliffs; and we must suppose that many a hidden stream beneath*

---

[1] A bradyseism is a slow quiet upward or downward movement of the earth's crust first observed in the Phlegraean Fields in southern Italy.

*the earth's back violently rolls its waves and submerged boulders; for the facts themselves demand that she be everywhere like herself. Since therefore she has these things attached beneath her and ranged beneath, the upper earth trembles under the shock of some great collapse when time undermines those huge caverns beneath; for whole mountains fall, and with the great shock the tremblings in an instant creep abroad from the place far and wide.*

### Appendix: Deduction of Eqs. (18) and (19) from Eq. (6)

We start from the weak-form of the force balance equations:

$$0 = \int_{\Omega_S} \mathbf{t}_s : \delta\mathbf{e}\, dV + \int_{\Omega_F} \mathbf{t}_F : \delta\mathbf{e}\, dV + \frac{d}{dt}\left(\int_{\Omega_S} \varrho_S \dot{\mathbf{u}} \cdot \delta\mathbf{u}\, dV + \int_{\Omega_F} \varrho_F \dot{\mathbf{u}} \cdot \delta\mathbf{u}\, dV\right)$$
$$- \int_{\Omega_S} \varrho_S \mathbf{b} \cdot \delta\mathbf{u}\, dV - \int_{\Omega_F} \varrho_F \mathbf{b} \cdot \delta\mathbf{u}\, dV. \tag{59}$$

Let us define $\forall \mathbf{x} \in \Omega$

$$\mathbf{s}(\mathbf{x}, t) := \begin{cases} \mathbf{t}_S(\mathbf{x}, t) & \forall \mathbf{x} \in \Omega_S \\ \mathbf{t}_F - p\mathbf{I} & \forall \mathbf{x} \in \Omega_F. \end{cases} \tag{60}$$

We obtain, by substituting for each term the expressions derived in (7) and (8), and by neglecting all terms of order equal to or higher than $\varepsilon$,

$$\int_\Omega \mathbf{s} : \delta\mathbf{e}\, dV \cong \sum_{\mathbf{x}\in\Lambda} \int_{V(\mathbf{x})} \mathbf{s} : \delta\mathbf{e}\, dV \cong \sum_{\mathbf{x}\in\Lambda} \int_{V(\mathbf{x})} [\mathbf{s}:\delta\mathbf{E} + \mathbf{s}:(\nabla h_I^A \otimes \delta\mathbf{Q}^A)]\, dV$$
$$= \sum_{\mathbf{x}\in\Lambda} [\langle\mathbf{s}\rangle_\mathbf{x} : \delta\mathbf{E} + \langle\mathbf{s}.\nabla h_I^A\rangle_\mathbf{x} .\delta\mathbf{Q}^A]|V| \cong \int_\Omega (\mathbf{S}:\delta\mathbf{E} + \mathbf{H}_I^A \cdot \delta\mathbf{Q}^A)\, dV, \tag{61}$$

where

$$\mathbf{S} = \langle\mathbf{C}\rangle : \mathbf{E} + \langle\mathbf{C}.\nabla h_I^A\rangle .\mathbf{Q}^A + \langle\mathbf{D}\rangle : \dot{\mathbf{E}} + \langle\mathbf{D}.\nabla h_I^A\rangle .\dot{\mathbf{Q}}^A, \tag{62}$$

$$\mathbf{H}^A = \langle\mathbf{C}.\nabla h_I^A\rangle : \mathbf{E} + \langle\mathbf{C}:(\nabla h_I^A \otimes \nabla h_I^B)\rangle .\mathbf{Q}^B + \langle\mathbf{D}.\nabla h_I^A\rangle : \dot{\mathbf{E}} + \langle\mathbf{D}:(\nabla h_I^A \otimes \nabla h_I^B)\rangle .\dot{\mathbf{Q}}^B, \tag{63}$$

$\forall t \in (t_0, t_f)$ and $\forall \mathbf{x} \in \Omega^o : \mathbf{S}(\mathbf{x}, t) := \langle\mathbf{s}\rangle_\mathbf{x}^S, \quad \mathbf{H}^A(\mathbf{x}, t) := \langle\mathbf{s}\nabla h_I^A\rangle_\mathbf{x}^S$.

In the same way we have

$$\int_{\Omega_F} \varrho \dot{\mathbf{u}} \cdot \delta\mathbf{u}\, dV + \int_{\Omega_S} \varrho \dot{\mathbf{u}} \cdot \delta\mathbf{u}\, dV$$
$$\cong \sum_{\mathbf{x}\in\Lambda} \int_{V_F(\mathbf{x})} \varrho(\dot{\mathbf{U}} + h_I^A \dot{\mathbf{Q}}^A) \cdot (\delta\mathbf{U} + h_I^B \delta\mathbf{Q}^B)\, dV + \sum_{\mathbf{x}\in\Lambda} \int_{V_S(\mathbf{x})} \varrho(\dot{\mathbf{U}} + h_I^A \dot{\mathbf{Q}}^A) \cdot (\delta\mathbf{U} + h_I^B \delta\mathbf{Q}^B)\, dV$$
$$\cong \int_\Omega (\langle\varrho\rangle \dot{\mathbf{U}} + \langle\varrho h_I^A\rangle \dot{\mathbf{Q}}^A) \cdot \delta\mathbf{U}\, dV + \int_\Omega (\langle\varrho h_I^A\rangle \dot{\mathbf{U}} + \langle\varrho h_I^A h_I^B\rangle \dot{\mathbf{Q}}^B) \cdot \delta\mathbf{Q}^A\, dV, \tag{64}$$

$$\int_{\Omega_F} \varrho\mathbf{b} \cdot \delta\mathbf{u}\, dV + \int_{\Omega_S} \varrho\mathbf{b} \cdot \delta\mathbf{u}\, dV$$
$$\cong \sum_{\mathbf{x}\in\Lambda} \int_{V_F(\mathbf{x})} [\varrho\mathbf{b} \cdot \delta\mathbf{U} + \varrho\mathbf{b} \cdot h_I^A \delta\dot{\mathbf{Q}}^A]\, dV + \sum_{\mathbf{x}\in\Lambda} \int_{V_S(\mathbf{x})} [\varrho\mathbf{b} \cdot \delta\mathbf{U} + \varrho\mathbf{b} \cdot h_I^A \delta\dot{\mathbf{Q}}^A]\, dV$$
$$\cong \int_\Omega [\langle\varrho\mathbf{b}\rangle \cdot \delta\mathbf{U} + \langle\varrho\mathbf{b} h_I^A\rangle \dot{\mathbf{Q}}^B) \cdot \delta\mathbf{Q}^A]\, dV, \tag{65}$$

$$\int_{\Omega_F} p \text{ div} (\delta\mathbf{u}) \, dV = \sum_{\mathbf{x} \in \Lambda} \int_{V_F(\mathbf{x})} p \text{ div} (\delta\mathbf{u}) \, dV = \sum_{\mathbf{x} \in \Lambda} \int_{V_F(\mathbf{x})} p \text{ div} (\delta\mathbf{U} + h^A \delta\mathbf{Q}^A) \, dV$$

$$= \sum_{\mathbf{x} \in \Lambda} \int_{V_F(\mathbf{x})} \{p \text{ div} (\delta\mathbf{U}) \, dV + p\nabla h^A \cdot \delta\mathbf{Q}^A \, dV + ph^A \text{ div} (\delta\mathbf{Q}^A) \, dV\}$$

$$= \sum_{\mathbf{x} \in \Lambda} \{\langle p \text{ div} (\delta\mathbf{U})\rangle_\mathbf{x}^F + \langle p\nabla h^A \rangle_\mathbf{x}^F \cdot \delta\mathbf{Q}^Q + \langle ph^A \rangle_\mathbf{x}^F \text{ div} (\delta\mathbf{Q}^A)\} |V|$$

$$\simeq \int_\Omega \{p \text{ div} (\delta\mathbf{U}) + \mathbf{P}^A \cdot \delta\mathbf{Q}^A + \langle ph^A \rangle \text{ div} (\delta\mathbf{Q}^A)\} \, dV$$

$$= p\delta\mathbf{U}|_{\partial\Omega} - \int_\Omega \nabla P \cdot \delta\mathbf{U} \, dV + \int_\Omega \mathbf{P}^A \cdot \delta\mathbf{Q}^A \, dV + \langle ph^A \rangle \delta\mathbf{Q}^A|_{\partial\Omega} - \int_\Omega (\nabla \langle ph^A \rangle) \delta\mathbf{Q}^A \, dV$$

$$= \int_\Omega (-\nabla P \cdot \delta\mathbf{U} + \mathbf{P}^A \cdot \delta\mathbf{Q}^A) \, dV \tag{66}$$

where we have defined

$$P = \langle p \rangle_\mathbf{x}^F, \quad \mathbf{P}^A = \langle p\nabla h_l^A \rangle_\mathbf{x}^F, \quad \forall \mathbf{x} \in \Omega^o.$$

Equation (65) holds because $\langle ph^A \rangle = 0$, Eqs. (60), (63), (64) and (65) hold $\forall \delta\mathbf{U}, \delta\mathbf{Q}^A : \delta\mathbf{U}|_{\partial\Omega} = 0$, $\delta\mathbf{Q}^A|_{\partial\Omega} = 0$.

Finally, we obtain

$$0 = \int_{\Omega_S} \mathbf{t}_s : \delta\mathbf{e} \, dV + \int_{\Omega_F} \mathbf{t}_F : \delta\mathbf{e} \, dV + \frac{d}{dt}\left(\int_{\Omega_S} \varrho_S \dot{\mathbf{u}} \cdot \delta\mathbf{u} \, dV + \int_{\Omega_F} \varrho_F \dot{\mathbf{u}} \cdot \delta\mathbf{u} \, dV\right)$$

$$- \int_{\Omega_S} \varrho_S \mathbf{b} \cdot \delta\mathbf{u} \, dV - \int_{\Omega_F} \varrho_F \mathbf{b} \cdot \delta\mathbf{u} \, dV$$

$$\cong \int_\Omega (\mathbf{S} : \delta\mathbf{E} + \mathbf{H}^A \cdot \delta\mathbf{Q}^A) \, dV + \int_\Omega (\langle \varrho \rangle \ddot{\mathbf{U}} + \langle \varrho h_l^A \rangle \ddot{\mathbf{Q}}^A) \cdot \delta\mathbf{U} \, dV$$

$$+ \int_\Omega (\langle \varrho h_l^A \rangle \ddot{\mathbf{U}} + \langle \varrho h_l^A h_l^B \rangle \ddot{\mathbf{Q}}^B) \cdot \delta\mathbf{Q}^A \, dV$$

$$- \int_\Omega [(\langle \varrho \mathbf{b} \rangle \cdot \delta\mathbf{U} + \langle \varrho \mathbf{b} h_l^A \rangle \cdot \delta\mathbf{Q}^A) + (\nabla P \cdot \delta\mathbf{U} + \mathbf{P}^A \cdot \delta\mathbf{Q}^A)] \, dV,$$

$$\forall \delta\mathbf{U}, \delta\mathbf{Q}^A : \delta\mathbf{U}|_{\partial\Omega} = 0, \quad \delta\mathbf{Q}^A|_{\partial\Omega} = 0, \tag{67}$$

from which

$$\text{div } \mathbf{S} - \langle \varrho \rangle \ddot{\mathbf{U}} - \langle \varrho h_l^A \rangle \ddot{\mathbf{Q}}^A + \nabla P + \langle \varrho \mathbf{b} \rangle = 0 \tag{68}$$

$$\langle \varrho h_l^A h_l^B \rangle \ddot{\mathbf{Q}}^B + \langle \varrho h_l^A \rangle \ddot{\mathbf{U}} + \mathbf{H}^A + \mathbf{P}^A - \langle \varrho \mathbf{b} h_l^A \rangle = 0. \tag{69}$$

**Authors' addresses:** F. dell'Isola, Dipartimento di Ingegneria Strutturale e Geotecnica, L. Rosa, Dottorato per la ricerca in Meccanica Teorica e Applicata, Università di Roma "La Sapienza", Via Eudossiana 18, Rome, Italy; C. Woźniak, Center of Mechanics IPPT PAN, Swietokrzyska 21, Warsaw, Poland